\documentclass[]{spie}  %>>> use for US letter paper
\usepackage{amsmath,amsfonts,amssymb}
\usepackage{graphicx}
\usepackage[colorlinks=true, allcolors=blue]{hyperref}
\usepackage{siunitx}
\usepackage{caption}
\newcommand{\fsky}{f_\textrm{sky}}
\def\sptthreeg{SPT-3G}
\def\sptnew{SPT-3G+}
\def\planck{\textit{Planck}}

\newcommand{\Deltazre}{\Delta z_{\rm re}}
\newcommand{\sigmadeltazreksz}{0.42}
\newcommand{\sigmatauksz}{0.005}

\DeclareSIUnit\lightspeed{$c$}
\DeclareSIUnit\rydberg{Ry}
\DeclareSIUnit\erg{erg}
\DeclareSIUnit\magnitude{mag}
\DeclareSIUnit\jansky{Jy}
\DeclareSIUnit\gauss{G}
\DeclareSIUnit\h{$h$}
\DeclareSIUnit\hseven{$h$_7}
\DeclareSIUnit\parsec{pc}
\DeclareSIUnit\year{yr}
\DeclareSIUnit\solarluminosity{\ensuremath{L_\sun}}
\DeclareSIUnit\solarmass{\ensuremath{M_\sun}}
\DeclareSIUnit\solarmassinenergy{\ensuremath{M_\sun|c^2}}
\DeclareSIUnit\solarradius{\ensuremath{R_\sun}}

\title{SPT-3G+: Mapping the High-Frequency Cosmic Microwave Background Using Kinetic Inductance Detectors}

\author[a,b,c]{A.~J.~Anderson}
\author[d]{P.~Barry}
\author[e,b,c]{A.~N.~Bender}
\author[a,b,c]{B.~A.~Benson}
\author[e,b]{L.~E.~Bleem}
\author[b,f,g,e,c]{J.~E.~Carlstrom}
\author[e]{T.~W.~Cecil}
\author[e,b,c]{C.~L.~Chang}
\author[b,c]{T.~M.~Crawford}
\author[c,b]{K.~R.~Dibert}
\author[h,i]{M.~A.~Dobbs}
\author[g]{K.~Fichman}
\author[j,k]{N.~W.~Halverson}
\author[l]{W.~L.~Holzapfel}
\author[g]{A.~Hryciuk}
\author[b,c]{K.~S.~Karkare}
\author[e]{J.~Li}
\author[e]{M.~Lisovenko}
\author[m]{D.~Marrone}
\author[b,c]{J.~McMahon}
\author[h]{J.~Montgomery}
\author[b]{T.~Natoli}
\author[e,b]{Z.~Pan}
\author[n]{S.~Raghunathan}
\author[o]{C.~L.~Reichardt}
\author[h]{M.~Rouble}
\author[b,c]{E.~Shirokoff}
\author[p]{G.~Smecher}
\author[q]{A.~A.~Stark}
\author[r,s,n]{J.~D.~Vieira}
\author[a,b]{M.~R.~Young}

\affil[a]{Fermi National Accelerator Laboratory, MS209, P.O. Box 500, Batavia, IL, 60510, USA}
\affil[b]{Kavli Institute for Cosmological Physics, University of Chicago, 5640 South Ellis Avenue, Chicago, IL, 60637, USA}
\affil[c]{Department of Astronomy and Astrophysics, University of Chicago, 5640 South Ellis Avenue, Chicago, IL, 60637, USA}
\affil[d]{School of Physics and Astronomy, Cardiff University, Cardiff CF24 3YB, United Kingdom}
\affil[e]{High-Energy Physics Division, Argonne National Laboratory, 9700 South Cass Avenue, Lemont, IL, 60439, USA}
\affil[f]{Enrico Fermi Institute, University of Chicago, 5640 South Ellis Avenue, Chicago, IL, 60637, USA}
\affil[g]{Department of Physics, University of Chicago, 5640 South Ellis Avenue, Chicago, IL, 60637, USA}
\affil[h]{Department of Physics and McGill Space Institute, McGill University, 3600 Rue University, Montreal, Quebec H3A 2T8, Canada}
\affil[i]{Canadian Institute for Advanced Research, CIFAR Program in Gravity and the Extreme Universe, Toronto, ON, M5G 1Z8, Canada}
\affil[j]{CASA, Department of Astrophysical and Planetary Sciences, University of Colorado, Boulder, CO, 80309, USA }
\affil[k]{Department of Physics, University of Colorado, Boulder, CO, 80309, USA}
\affil[l]{Department of Physics, University of California, Berkeley, CA, 94720, USA}
\affil[m]{Steward Observatory and Department of Astronomy, University of Arizona, 933 North Cherry Avenue, Tucson, AZ 85721, USA}
\affil[n]{Center for AstroPhysical Surveys, National Center for Supercomputing Applications, Urbana, IL, 61801, USA}
\affil[o]{School of Physics, University of Melbourne, Parkville, VIC 3010, Australia}
\affil[p]{Three-Speed Logic, Inc., Victoria, B.C., V8S 3Z5, Canada}
\affil[q]{Harvard-Smithsonian Center for Astrophysics, 60 Garden Street, Cambridge, MA, 02138, USA}
\affil[r]{Department of Astronomy, University of Illinois Urbana-Champaign, 1002 West Green Street, Urbana, IL, 61801, USA}
\affil[s]{Department of Physics, University of Illinois Urbana-Champaign, 1110 West Green Street, Urbana, IL, 61801, USA}

\begin{document} 
\maketitle

\begin{abstract}
We present the design and science goals of SPT-3G+, a new camera for the South Pole Telescope, which will consist of a dense array of \num{34100} kinetic inductance detectors measuring the cosmic microwave background (CMB) at \SIlist[list-units=single]{220;285;345}{\giga\hertz}.
%as well as spectroscopy of millimeter-wavelength emission from point sources.
The SPT-3G+ dataset will enable new constraints on the process of reionization, including measurements of the patchy kinematic Sunyaev-Zeldovich effect and improved constraints on the optical depth due to reionization.
At the same time, it will serve as a pathfinder for the detection of Rayleigh scattering, which could allow future CMB surveys to constrain cosmological parameters better than from the primary CMB alone.
In addition, the combined, multi-band SPT-3G and SPT-3G+ survey data, will have several synergies that enhance the original SPT-3G survey, including: extending the redshift-reach of SZ cluster surveys to $z > 2$; understanding the relationship between magnetic fields and star formation in our Galaxy; improved characterization of the impact of dust on inflationary B-mode searches; and characterizing astrophysical transients at the boundary between mm and sub-mm wavelengths.
Finally, the modular design of the SPT-3G+ camera allows it to serve as an on-sky demonstrator for new detector technologies employing microwave readout, such as the on-chip spectrometers that we expect to deploy during the SPT-3G+ survey.
In this paper, we describe the science goals of the project and the key technology developments that enable its powerful yet compact design.
\end{abstract}

\keywords{cosmic microwave background, kinetic inductance detectors, South Pole Telescope, kinematic Sunyaev-Zel'dovich effect}

\section{Introduction}
\label{sec:intro} 

A combination of satellite and ground-based experiments measuring the temperature and polarization of the cosmic microwave background (CMB) have produced sub-percent constraints on $\Lambda$CDM cosmological parameters.
Ground-based experiments such as the existing South Pole Observatory~\cite{hui18,Sobrin2021} and the upcoming Simons Observatory~\cite{simonsobservatorycollab19} and CMB-S4~\cite{cmbs4collab19} will continue to advance these measurements of the primary CMB.
Meanwhile, measurements of secondary CMB anisotropies, produced by the interactions of CMB photons with matter along the line of sight after the epoch of recombination, yield science that is highly complementary to that of the primary CMB.
Measurements of the thermal and kinematic Sunyaev-Zeldovich effects (tSZ, kSZ), for example, can be used to improve our understanding of galaxy cluster formation, as well as to infer both the duration of the epoch of reionization and the associated optical depth, which can improve constraints on other cosmological parameters such as the sum of the neutrino masses~\cite{abazajian16}.
%as well as to infer the optical depth due to reionization and its duration, which can improve constraints on other cosmological parameters such as the sum of the neutrino masses~\cite{abazajian16}.
Improving measurements of secondary anisotropies requires observations at higher, sub-mm frequencies and sub-arcminute angular resolution, in order to better characterize and remove astrophysical foregrounds, including the cosmic infrared background (CIB).
%CMB cameras optimized for measuring these secondary anisotropies possess different characteristics than those optimized for measuring the primary CMB, including sub-arcminute angular resolution and measurements at higher, sub-mm frequencies.
A set of upcoming projects, including PrimeCam on the CCAT-prime telescope~\cite{CCATp2021} and our new camera, SPT-3G+, for the 10-m South Pole Telescope (SPT), are being developed to conduct surveys optimized for detecting CMB secondaries.
%A CMB camera optimized for detecting secondary anisotropies need not share the same design as one optimized to study the primary CMB.
%Indeed, the 10-m South Pole Telescope (SPT) with its program of ultra-deep, arcminute- to subarcminute-resolution surveys is an ideal instrument for studying CMB secondaries.

SPT-3G+ is a powerful new CMB camera for the SPT that will enable precision measurements of these secondary CMB anisotropies by providing a combination of low noise, broad frequency coverage for excellent control of foregrounds, and sub-arcminute angular resolution.
SPT-3G+ will replace the currently operating camera, SPT-3G, with one having \num{34100} kinetic inductance detectors (KIDs) operating in three frequency bands centered at \SIlist[list-units=single]{220;285;345}{\giga\hertz}, while reusing the same ambient temperature optics.
Over the course of a 4-year survey, SPT-3G+ will observe the same \SI{1500}{deg^2} survey field as SPT-3G, with the combined dataset having noise levels of \SIlist[list-units=single]{3;2;3;6;28}{\micro\kelvin\textrm{-arcmin}} spanning 5 frequency bands centered at \SIlist[list-units=single]{95;150;220;285;345}{\giga\hertz}, with the excellent angular resolution (0.7~arcmin at 280~GHz) provided by the SPT primary mirror.
%By continuing to observe the same \SI{1500}{deg^2} survey field as SPT-3G, the combined SPT-3G and SPT-3G+ datasets will achieve noise levels of \SIlist[list-units=single]{3;2;3;6;28}{\micro\kelvin\textrm{-arcmin}} spanning 5 frequency bands centered at \SIlist[list-units=single]{95;150;220;285;345}{\giga\hertz}, with excellent angular resolution (0.7~arcmin at 280~GHz) provided by the SPT primary mirror.
The uniquely low atmospheric noise at the South Pole---one of the best developed sites for mm- and submm-wavelength observations---will enable the SPT-3G+ survey to study CMB secondaries across a range of angular scales, from the SZ effects at arcminute scales to Rayleigh scattering at sub-degree scales.
%Deployed at one of the best developed sites for mm- and submm-wavelength observations, the SPT-3G+ survey will benefit from the uniquely stable South Pole atmosphere
%permitting the use of degree-scale modes in polarization maps and only slightly smaller scales in temperature maps.

\section{Science Targets}
\label{sec:science}
The science program of SPT-3G+ focuses on three main themes: the physics of reionization, Rayleigh scattering, and the evolution of galaxies and clusters over cosmic time.
These science themes are studied with multiple observational probes, described in detail in the following sections.
The SPT-3G+ dataset will precisely constrain the optical depth due to reionization and its duration with its ultra-sensitive measurements of the kSZ effect (\autoref{sec:kSZ}), in addition to potentially probing the process of star formation during reionization with its mm- and submm-wavelength line-intensity mapping (LIM) measurements (\autoref{sec:lim}).
The combined data from SPT-3G and SPT-3G+ will serve as a pathfinder for future measurements of Rayleigh scattering that will reduce the effect of cosmic variance in cosmological parameter extraction (\autoref{sec:rayleigh}).
And the combination of ultra-deep maps and high-frequency observing bands will allow SPT-3G+ to study the assembly of proto-clusters into galaxy clusters at high redshift (\autoref{sec:clusters}) as well as the physics connecting magnetic fields and dust in our own local Galaxy (\autoref{sec:galacticdust}).

In addition, a fourth theme of SPT-3G+ is technology development.
New detector and readout technologies that are demonstrated successfully in the lab often require significant additional R\&D to succeed in the more demanding conditions presented by on-sky observing.
Translating technologies from the lab to on-sky conditions requires telescope and cryogenic platforms that are flexible enough to rapidly deploy new detectors, but which also provide dedicated access to long integration times to study instrumental systematics.
Following the lineage of the SPT-SZ~\cite{Chang2009}, SPTpol~\cite{Austermann2012}, and SPT-3G cameras~\cite{Sobrin2021}, which each demonstrated multiple early-stage technologies, SPT-3G+ provides an ideal platform for technology development, focusing specifically on KIDs for CMB observations and mm-wave LIM.

\subsection{Kinematic Sunyaev-Zeldovich Effect}
\label{sec:kSZ}
The kSZ effect is the anisotropy induced by CMB photons scattering off electrons with bulk peculiar velocities relative to the Hubble flow, and its correlation with the cosmic velocity field and ionization history can be used to constrain cosmological parameters.
The anisotropy is sourced at two distinct redshifts due to different processes, known as the \emph{reionization} or \emph{patchy kSZ} and the \emph{late-time kSZ} effects.
%The late-time kSZ effect arises due to the large-scale velocity of massive halos primarily during $0 \lesssim z \lesssim 3$, and it provides cosmological constraints through its correlation with the cosmic velocity field.
The late-time kSZ effect arises due to scattering of CMB photons off electrons in massive halos with a bulk velocity along the line of sight, primarily during $0 \lesssim z \lesssim 3$.
In particular, the late-time kSZ can be used to measure the growth of structure and therefore test models of dark energy and gravity on cosmic distance scales~\cite{Mueller2015, keisler13}.
On the other hand, the patchy kSZ effect arises due to scattering of CMB photons off expanding bubbles of free electrons with a bulk velocity along the line of sight, during reionization at $z \gtrsim 6$.
The optical depth due to reionization and its duration, which can be measured with the patchy kSZ effect~\cite{Alvarez2021}, are currently poorly constrained~\cite{reichardt20, planck18-6} but of great importance for extracting cosmological parameters from CMB data.
Constraints on the sum of the neutrino masses by the next generation of CMB experiments like CMB-S4 will be limited by the parameter degeneracy with the optical depth $\tau$; reducing the uncertainty on $\tau$ to the cosmic variance limit would improve the precision of the neutrino mass measurement by $\sim 40\%$~\cite{abazajian16}.

\begin{figure}[]
\centering
\includegraphics[width=0.7\textwidth]{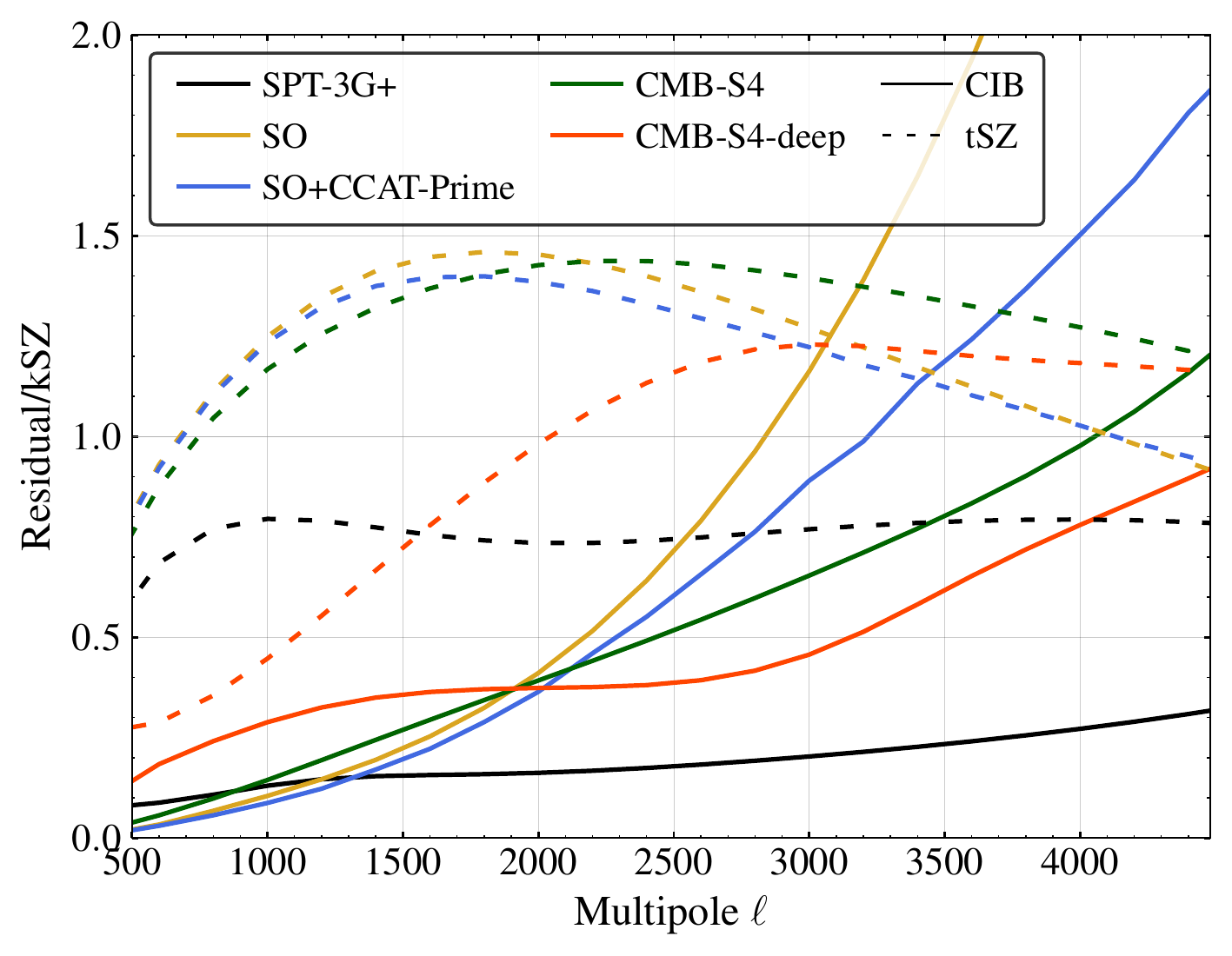}
\caption{
Ratio of residual CIB and tSZ contamination to kSZ power in a forecast of ILC spectra for several upcoming experiments.
Solid lines are residuals due to CIB, while dashed lines are residuals due to tSZ. 
The SPT-3G+ curves include \SI{90}{\giga\hertz} and \SI{150}{\giga\hertz} data from the currently operating SPT-3G survey.
The level of residual kSZ contamination is significantly lower for SPT-3G+ than other experiments, especially at high multipoles.
}
\label{fig:ilcresiduals}
\end{figure}

The primary experimental challenge in measuring the kSZ power spectrum is the separation of the signal from small-scale foreground contamination such as the cosmic infrared background (CIB) and tSZ effect.
Residual noise from these foregrounds in maps cleaned with multi-frequency methods such as internal linear combination (ILC) becomes larger than the kSZ signal itself at sufficiently high multipoles, and this practically limits the number of kSZ modes that can be used in cosmological analyses.
SPT-3G+ addresses this problem with its three high-frequency bands, which produce deep maps that can be combined directly with the lower-frequency SPT-3G maps on the same patch of the sky.
The resulting residuals due to both CIB and tSZ contamination, shown in \autoref{fig:ilcresiduals}, are significantly lower than for other upcoming experiments, allowing the use of higher-$\ell$ data with less sensitivity to foreground modeling.

\begin{figure}[]
\centering
\includegraphics[width=0.8\textwidth]{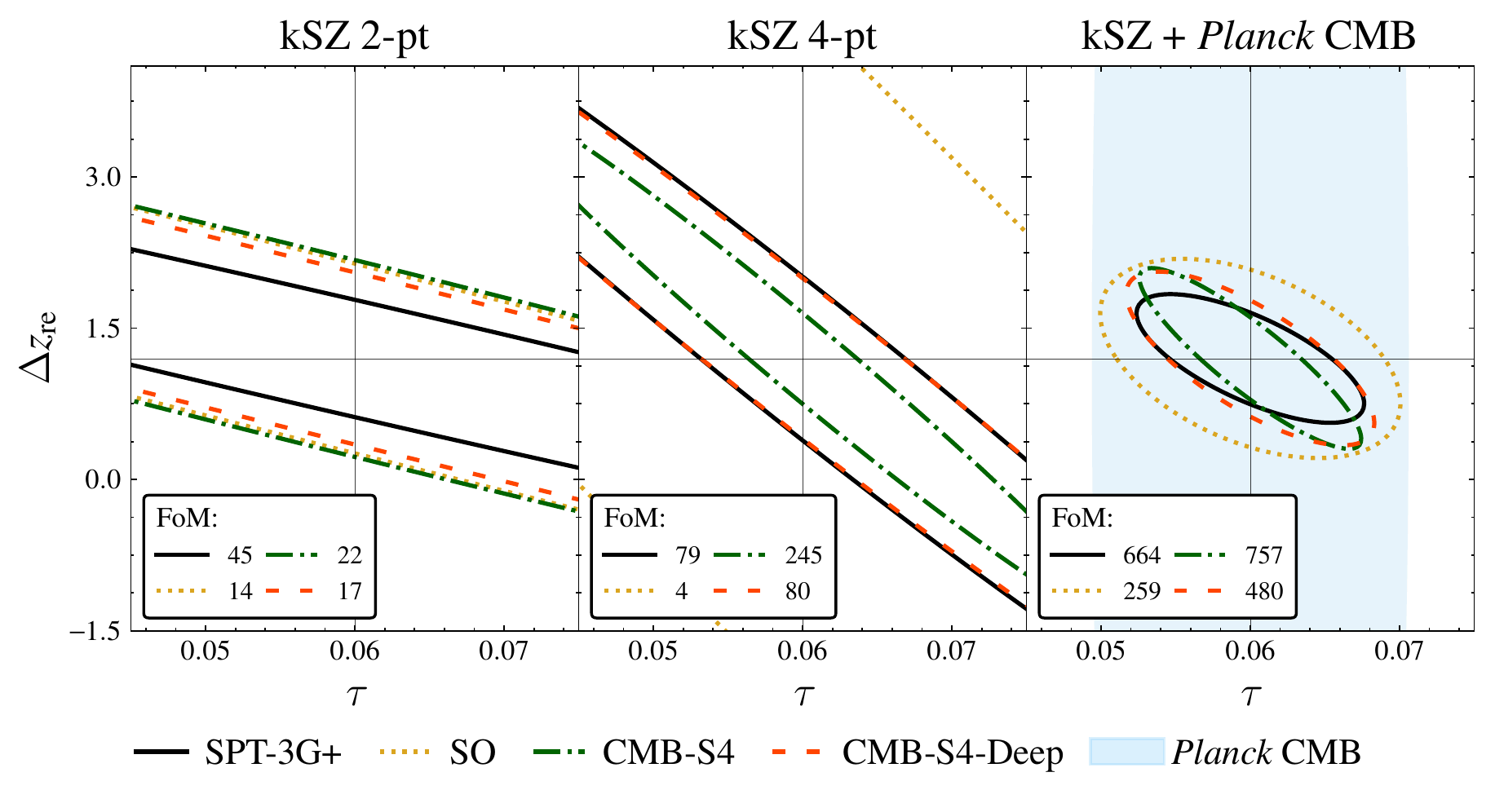}
\caption{
Constraints on the duration of reionization $\Delta z_\textrm{re}$ and the optical depth due to reionization $\tau$ from the kSZ power spectrum \emph{(left)}, the kSZ 4-point function \emph{(middle)}, and the combination of the two \emph{(right)}.
We define a figure of merit $\textrm{FoM} = 1/\sqrt{\det( \mathcal{F}^{-1})}$, with $\mathcal{F}$ the Fisher matrix, to quantify the statistical precision of the constraints.
SPT-3G+ will dramatically improve constraints on these reionization parameters, with a precision similar to CMB-S4 despite finishing its operations a decade earlier.
}
\label{fig:ksztau}
\end{figure}

The high-precision measurements of the kSZ effect by SPT-3G+ will enable powerful new constraints on $\tau$, thereby reducing the impact of parameter degeneracies in future surveys.
The patchy kSZ effect has a significant non-gaussian component due to the fact that variation in the cosmic velocity field along different lines of sight causes a modulation in the amplitude of the kSZ power spectrum across the sky.
This results in a nonzero kSZ 4-point function, the amplitude of which has a different dependence on the reionization parameters $\Delta z_\textrm{re}$ and $\tau$ than the kSZ power spectrum / 2-point function~\cite{Smith2016,Ferraro2018}.
The combination of the kSZ 2-point and 4-point functions can therefore provide a constraint on $\tau$ comparable to that from \planck\ low-$\ell$ E-mode data~\cite{Alvarez2021}.

%, as shown in \autoref{fig:ksztau}
We adopt a framework similar to previous studies~\cite{Ferraro2018,Alvarez2021} to forecast the constraints on reionization parameters expected from SPT-3G+, summarized in \autoref{fig:ksztau}.
Contamination from the CIB and our limited ability to model it precludes using data on scales smaller than a given $\ell_\textrm{max}$ for the kSZ measurement.
In our forecasts, we set $\ell_\textrm{max} = 4000$ for TT and $\ell_\textrm{max} = 5000$ for TE and EE.
The 4-point function is expected to be less susceptible to foreground contamination, so we choose $\ell_\textrm{max} = 7000$ for those data.
For SPT-3G+, we obtain $\sigma(\tau) = \sigmatauksz$, which is 20\% tighter than current constraints from the primary CMB measurements \cite{planck18-6}.
Even without the \planck\ data, the kSZ-only constraints are within 15\% of the \planck-only constraints, but achieved with a completely independent method, providing an important systematic check.
For the duration of reionization, we obtain $\sigma(\Deltazre) = \sigmadeltazreksz$.

\subsection{Rayleigh Scattering}
\label{sec:rayleigh}

\begin{figure}[]
\centering
\includegraphics[width=0.7\textwidth]{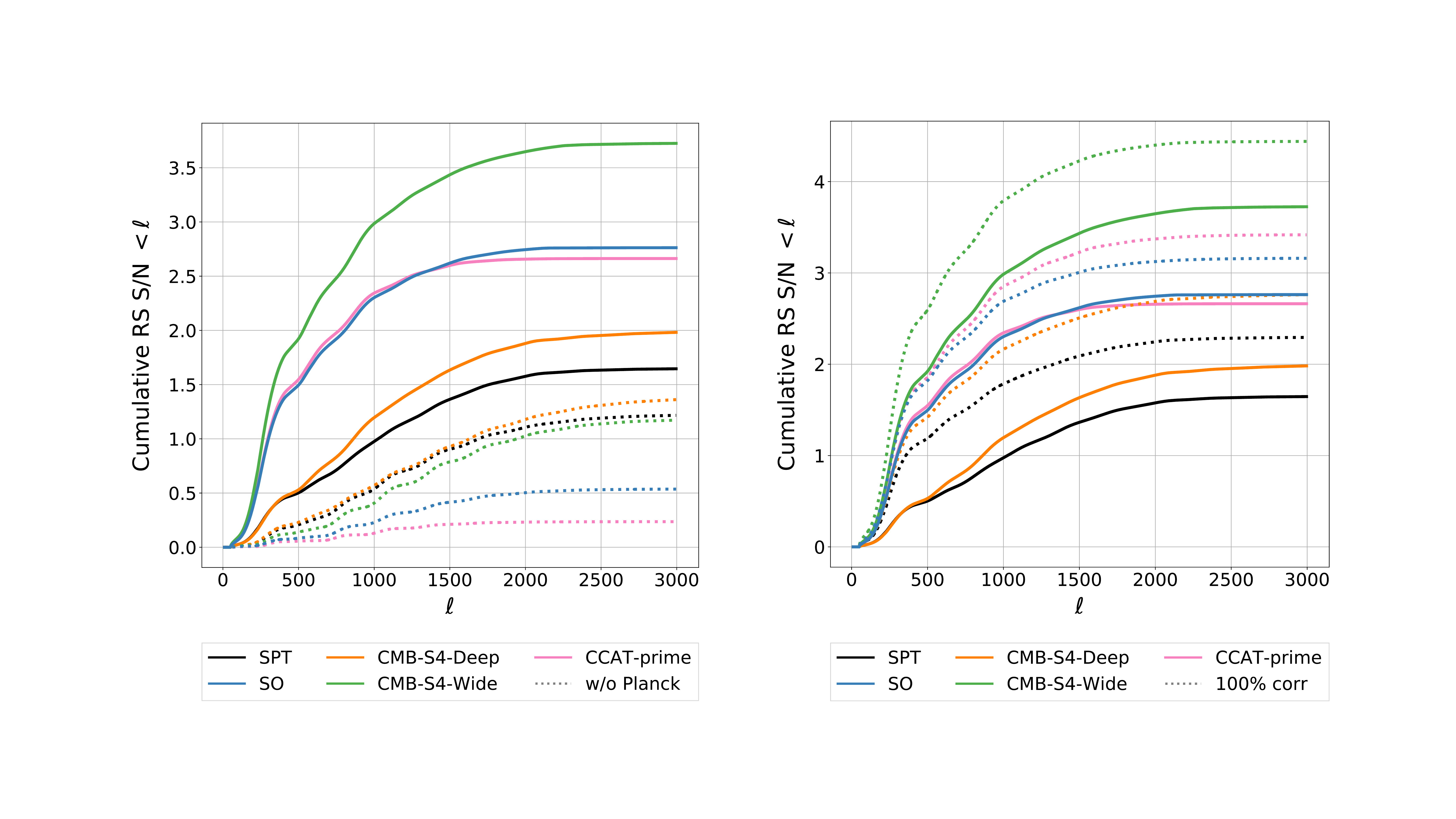}
\caption{
%Cumulative total signal-to-noise ratio for the RS signal using modes below a maximum multipole $\ell$ for several upcoming ground-based CMB experiments.
%Forecast includes the effects of atmospheric noise, detector noise, and galactic and extragalactic foregrounds.
%Solid lines include \planck\ data, while dashed lines do not.
%Deep surveys with small sky fraction, such as SPT-3G+ and CMB-S4-Deep (both $\fsky = 0.03$) contain more independent information from \planck\ than wide-field surveys, such as CCAT-prime and SO.
%Reproduced from Dibert, \emph{et al.}~\cite{Dibert2022}.
Cumulative total forecasted signal-to-noise ratio for the RS signal using modes below a maximum multipole $\ell$ for several upcoming ground-based CMB experiments.
Forecasts include the effects of atmospheric noise, detector noise, and galactic and extragalactic foregrounds.
Solid lines represent the combination of a given ground-based instrument with the \planck\ data corresponding to that instrument's observation patch on the sky.
Dotted lines represent forecasts for ground-based experiments alone.
Deep surveys with small sky fraction, such as SPT-3G+ and CMB-S4-Deep (both $\fsky = 0.03$)  contain more independent information from \planck\ than wide-field surveys, such as CCAT-prime and SO.
Reproduced from Dibert, \emph{et al.}~\cite{Dibert2022}.}
\label{fig:rayleighscattering}
\end{figure}

The Rayleigh scattering (RS) of CMB photons on neutral hydrogen atoms occurs shortly after recombination.
This process generates a secondary CMB anisotropy originating from an additional scattering surface, which has a frequency-dependent redshift slightly lower than that of the surface of last scattering.
Because the RS cross section is proportional to $\nu^4$, it produces a frequency-dependent distortion to the primary CMB temperature and polarization anisotropies, the amplitude of which is proportionally larger at higher frequencies~\cite{Lewis2013}.
Although its prediction relies on well understood physics, a first detection of RS would be a consistency test of the description of recombination in $\Lambda$CDM, and it would pave the way for using RS to improve measurements of cosmological parameters.
For instance, RS provides an independent constraint on the primordial helium fraction $Y_p$, breaking the degeneracy with $N_\textrm{eff}$ that is present in the primary CMB data alone~\cite{Alipour2015}; and future CMB space missions such as PICO may improve their constraints on the neutrino mass scale $\sum m_\nu$ by as much as $2\times$ by including RS~\cite{Beringue2021}.

SPT-3G+ has several features that are particularly well suited for overcoming the extreme challenge of detecting CMB RS from the ground.
First, the $\nu^4$ frequency scaling of RS means that the signal in the CMB is higher at the frequencies observed by SPT-3G+, above the \SI{150}{\giga\hertz} peak of the CMB blackbody.
Second, most of the signal-to-noise in the RS measurement comes from temperature, rather than polarization, at multipoles $\ell < 1000$, so atmospheric noise significantly limits sensitivity.
As discussed in detail in \autoref{sec:sitetelescope}, the South Pole atmosphere has consistently low, stable, sky noise for the entire austral winter, and is thus the ideal site at which to attempt the RS measurement.
Finally, the broad frequency coverage of SPT-3G+, in combination with SPT-3G data, enables excellent mitigation of extragalactic foregrounds, such as the CIB, which are the dominant source of noise in the temperature-based measurement of RS.

Forecasts of the RS signal-to-noise ratio for several upcoming ground-based experiments were performed in Dibert, \emph{et al.}~\cite{Dibert2022} and are presented in \autoref{fig:rayleighscattering} (see also similar results in Zhu, \emph{et al.}~\cite{Zhu2022}).
These forecasts include the effects of atmospheric noise, detector noise, and galactic and extragalactic foregrounds, and they combine with \planck{} data on the patch of the sky observed by each experiment.
The combined data of SPT-3G and SPT-3G+ will have an RS signal-to-noise of 1.6 to 2.3, depending on the degree of correlation of the atmospheric noise between different frequency bands.
While this is similar to the upcoming near-term experiments SO and CCAT-prime, if these two experiments are compared to SPT-3G+ without the addition of \planck\ data (dashed lines of \autoref{fig:rayleighscattering}), SPT-3G+ has the greatest RS signal-to-noise on its own.
This occurs because SO and CCAT-prime have much larger sky fractions than SPT-3G+ ($f_\textrm{sky} \sim 0.4$ vs. $f_\textrm{sky} \sim 0.03$), so the \planck\ data contributes relatively more statistical power to forecasts for those instruments.

\subsection{Delensing and Primordial B-modes}
\label{sec:delensing}
One of the key predictions of the inflationary paradigm is the production of gravitational waves that give rise to a B-mode polarization of the CMB on degree angular scales.
Multiple ongoing and future CMB experiments, including the BICEP/\emph{Keck} program~\cite{hui18}, Simons Observatory~\cite{simonsobservatorycollab19}, and CMB-S4~\cite{cmbs4collab19} are undertaking searches for degree-scale primordial B modes, the amplitude of which is proportional to the tensor-to-scalar ratio $r$.
Although low-resolution refracting telescopes have thus far produced the best upper limits on the value of $r$ due to their low cost and simplicity~\cite{bicep2keck21}, their sensitivity is limited by their inability to distinguish primordial B modes from those produced by the gravitational lensing of E modes by large-scale structure.
This motivates the use of high-resolution SPT maps to ``delens'' the low-resolution B-mode maps from BICEP Array~\cite{bkdelensing20}.

Motivated by the need for delensing to improve constraints on $r$, the SPT-3G and BICEP Array projects have formed an umbrella collaboration called the South Pole Observatory (SPO), in order to facilitate joint observations and analyses.
The \sptnew\ dataset will expand the ability of the SPT to delens BICEP Array data beyond the level of \sptthreeg. 
For example, \sptnew\ will enable new systematic checks for dust from the improved high-frequency sensitivity, with the \sptnew\ \SI{220}{\giga\hertz} band alone having a noise level comparable to the \sptthreeg\ \SI{90}{\giga\hertz} band, 
and, compared to \sptthreeg\ alone, \sptnew\ will increase the total survey weight by nearly 50\%.

\subsection{Growth of Galaxies and Clusters}
\label{sec:clusters}
Galaxy clusters are the largest gravitationally bound objects in the universe.
However, the details of how these structures are assembled are far from understood, and in particular, we do not yet know how the cluster environment affects galaxy evolution and the impact of changing star-formation rates~\cite{magliocchetti13, miller15}. 
%what roles may be played by star-formation downsizing~\cite{magliocchetti13, miller15}. 
Over a period of 2~Gyr (between $z \sim 2$ and $z \sim 4$), there must be a dramatic transformation between the proto-cluster stage, with dense aggregations of rapidly star forming galaxies, to the cluster stage, in which baryonic matter is divided between passive galaxies and a more massive halo of hot intracluster gas.
The combination of \sptthreeg\ with the higher angular resolution and improved dust sensitivity of \sptnew\ will allow us to explore this frontier, detecting the intracluster medium in clusters at $z>2$ through the tSZ effect, even in the presence of correlated dust emission. 
Added to \sptthreeg\ maps, \sptnew\ data will increase the yield of high redshift clusters by a factor of $\gtrsim 2$, discovering $\sim 200$ massive SZ clusters at $z>2$.

% may not want to introduce SPT-SZ here, since it hasn't been mentioned in detail before this
\sptnew\ will provide an unprecedented submm-wave survey and catalog to the extragalactic community.
The extended frequency coverage and higher sensitivity will yield 1000$\times$ more dusty sources at $z>1$ than SPT-SZ; whereas SPT-SZ discovered one source at $z=6.9$ \cite{marrone18}, \sptnew\ will discover more than 200 at $z>7$.
This will complement the contemporaneous survey of CCAT-prime, which will observe over \SI{200}{deg^2} to the confusion limit, detecting \SI{1200} galaxies at $5<z<8$~\cite{CCATp2021}.
The higher signal-to-noise ratio and smaller beam area from \sptnew\ will narrow the uncertainty region for discovered sources, greatly improving our ability to associate them with counterparts at other wavelengths. 

The combination of the SPT-3G and \sptnew\ surveys will provide a unique and powerful catalog of sources that will be crucial to the galaxy evolution community in the coming decade. \sptnew\ will have strong synergies with major US facilities such as ALMA and JWST by providing an ultra-wide survey field for identifying the rarest and most interesting objects to study in great detail, and optical and infrared surveys such as Rubin/LSST and \textit{Roman} by uncovering the objects otherwise hidden by dust.

\subsection{Galactic Dust and Magnetic Fields}
\label{sec:galacticdust}
The deep, polarization-sensitive \sptnew\ data will provide a detailed view of the diffuse interstellar medium (ISM) in our Galaxy and help elucidate the role magnetic fields play in star-forming molecular clouds.
We know this role is important: the energy density of the ordered and turbulent magnetic fields is comparable to the turbulent energy of the ISM \cite{planck15-19}.
While existing facilities have been used to reconstruct the magnetic field in individual clouds~\cite{guerra20}, the \sptnew\ Galactic survey will measure the linear polarization of Galactic dust in a large sample of nearby molecular clouds with resolution 10$\times$ finer than \planck\, down to subparsec scales, enabling robust statistical inference of the role magnetic fields play in the star-formation process.
Along with CCAT-prime~\cite{CCATp2021}, \sptnew\ will provide one of the first large-area submm-wave Galactic surveys with sufficient angular resolution to measure the magnetic field structure in molecular clouds at 0.1\,pc resolution for clouds within 680 pc, and at 1\,pc resolution for clouds within 6.8\,kpc.
The former would allow detailed studies of local molecular clouds down to the filament scale, while the latter would yield a large statistical sample of $\sim$1300 molecular clouds, enabling a detailed statistical exploration of the connection between magnetic fields and star formation as a function of cloud properties.

\sptnew\ will also improve on recent measurements of polarized dust emission by \planck\ that have been used to make significant advances in the understanding of magnetically induced dust anisotropy \cite{clark15}, ISM turbulence \cite{caldwell17}, and the Galactic magnetic field \cite{bracco19}.
The \sptnew\ measurements would be an order of magnitude more sensitive to polarized dust emission than the \planck\ data in this region of sky and frequency range and will significantly improve our understanding of magnetized ISM and the polarized foregrounds that are key to the search for gravitational-wave signatures in the CMB \cite{bicep2keck18, cmbs4collab19}. 

\subsection{Astrophysical Transients}
\label{sec:transients}
With its high instantaneous sensitivity and an observation cadence that revisits the same patches of the sky at intervals ranging from a few hours to a few days, SPT-3G+ will detect a wide array of astrophysical transients.
% highlight that this is open science / discovery space with sentence here
In recent years, SPTpol and SPT-3G have pioneered the study of mm-wave transients, making the first potential detection by a CMB experiment of a gamma-ray burst (GRB) afterglow~\cite{whitehorn16}, initiating a dedicated survey for mm-wave transients that has detected emission from flaring stars and possible extragalactic sources~\cite{guns21}, and performing the first measurements of asteroids with a dedicated CMB survey~\cite{Chichura2022}.

\sptnew\ will have similar overall raw flux sensitivity to SPT-3G, but at higher frequencies, with a sensitivity to transients of 1-day duration of \SIlist[list-units=single]{4.6;4.5;20}{\milli\jansky} in the \SIlist[list-units=single]{220;285;345}{\giga\hertz} bands respectively.
Notably, the flux sensitivity for \sptnew\ is more than $5\times$ better at \SI{220}{\giga\hertz} than SPT-3G.
Since the spectrum of GRBs is expected to be relatively flat at millimeter-wavelengths~\cite{granot02a}, \sptnew\ should detect a similar number of GRB sources as \sptthreeg\, but in a higher frequency range where no dedicated surveys for transient sources have ever been performed.
This will enable tests of models of the spectral energy distributions of GRB afterglows.
In addition, \sptnew\ will be several times more sensitive than \sptthreeg\ to asteroids due to their thermal spectrum that is brighter at higher frequencies.
These mm-wave measurements are sensitive to the properties of asteroid composition several wavelengths below the surface and therefore can constrain models of the temperature and wavelength-dependent emissivity of asteroid regoliths~\cite{keihm13}.

\subsection{Line-Intensity Mapping}
\label{sec:lim}
Line-intensity mapping (LIM) is a powerful, but relatively new, observational technique which probes a wide range of redshifts by mapping the intensity of atomic and molecular emission lines both spatially and spectrally.
Far-IR emission lines such as the rotational transitions of CO or the [CII] fine structure line emitted over $0 \lesssim z \lesssim 10$ are redshifted into the same atmospheric frequency ``windows'' at which ground-based CMB experiments such as SPT-3G and SPT-3G+ observe, meaning that LIM observations can be performed with detectors and optics similar to those used by CMB experiments.
There is an extensive literature exploring science opportunities from mm- and submm-wavelength LIM observations (e.g. see Kovetz, \emph{et al.}~\cite{Kovetz2019}).
For example, at $z \lesssim 6$, future large-scale LIM surveys probe large-scale structure and therefore can constrain fundamental cosmological parameters such as the sum of the neutrino masses $\sum m_\nu$~\cite{MoradinezhadDizgah2022} and the amplitude of non-gaussianity in the primordial curvature perturbations, parameterized by $f_\textrm{NL}$~\cite{MoradinezhadDizgah2018}.
The latter is a powerful probe of inflation, which can distinguish between single- and multi-field inflationary models.
At higher redshifts, observations of [CII] would provide a probe of the epoch of reionization and would constrain the star formation rate of the earliest galaxies~\cite{Sun2021}.

SPT-3G+ will provide a platform for LIM observations capable of fielding on the order of 1000 spectroscopic pixels with $R = \lambda / \Delta \lambda \sim 100 - 300$, with a staged deployment of detectors in the latter half of the SPT-3G+ survey.
Combined with the $\gtrsim 90\%$ observing efficiency of the SPT-3G+ camera, such a LIM survey would represent an increase of several orders of magnitude in survey depth (measured in number of spectrometers~$\times$~observation time) compared with ongoing projects~\cite{KarkareSLIMLTD2022}.
The on-chip spectrometer technology described in \autoref{sec:limdetectors} is fully compatible with the optics and readout electronics planned for SPT-3G+, enabling it to seamlessly transition between CMB and LIM survey modes.

\section{Survey}
\label{sec:survey}
To achieve these science goals, SPT-3G+ will carry out two surveys: a ``Main'' survey targeting reionization, Rayleigh scattering, and high-redshift galaxy clusters; and a ``Galactic'' survey targeting the measurements of dust in our Galaxy.
The Main survey will cover the same \SI{1500}{deg^2} footprint observed by the ongoing SPT-3G experiment, which will produce some of the deepest arcminute-resolution CMB maps upon its completion, with map noise levels of \SIlist{3;2}{\micro\kelvin\textrm{-}arcmin} in temperature at \SIlist{95;150}{\giga\hertz}.
Replicating the cadence of SPT-3G, the new camera will observe this footprint each year from late March through the beginning of December, corresponding to the months with the best atmospheric conditions at South Pole.
The combined ultra-low-noise maps in five frequencies provide the stringent control of CIB and tSZ residuals that enable the kSZ measurements described in \autoref{sec:kSZ}. 

During the austral summer months from December through late March, SPT-3G+ will observe a new \SI{7000}{deg^2} footprint covering much of the galaxy.
In this part of the year, the diffraction sidelobes of the SPT primary mirror intercept the sun when the telescope is pointed at the Main field, resulting in significant spurious features in temperature maps.
This motivates observing the opposite half of the southern sky to a shallower depth, which will be used for the Galactic science theme of SPT-3G+.

\begin{table}[]
\def\arraystretch{1.0}
\setlength{\tabcolsep}{7pt}
\centering
\begin{tabular}{l  c  c  c  c}
\hline\hline
Survey & Area & 220\,GHz T noise & 285\,GHz T noise & 345\,GHz T noise \\
  & [\si{deg^2}] & [\si{\micro\kelvin\textrm{-}arcmin}] & [\si{\micro\kelvin\textrm{-}arcmin}] & [\si{\micro\kelvin\textrm{-}arcmin}] \\
\hline
Main & \num{1500} & 2.9 & 5.6 & 28 \\
Galactic & \num{7000} & 13 & 25 & 130 \\
\hline
\end{tabular}
\caption{
Noise levels of the two surveys to be performed by SPT-3G+, assuming four years of operation.
The Main survey covers the same \SI{1500}{deg^2} field currently being observed by the SPT-3G camera in frequency bands centered at \SIlist{95;150;220}{\giga\hertz} and will be conducted during the austral winter for approximately 9 months of the year.
The Galactic survey will observe a \SI{7000}{deg^2} area to a shallower depth, using the approximately 3 months of the year when the Main field is contaminated by sidelobes from the Sun.
}
\label{tab:surveys}
\end{table}

\begin{figure}[]
\centering
\includegraphics[width=0.8\textwidth]{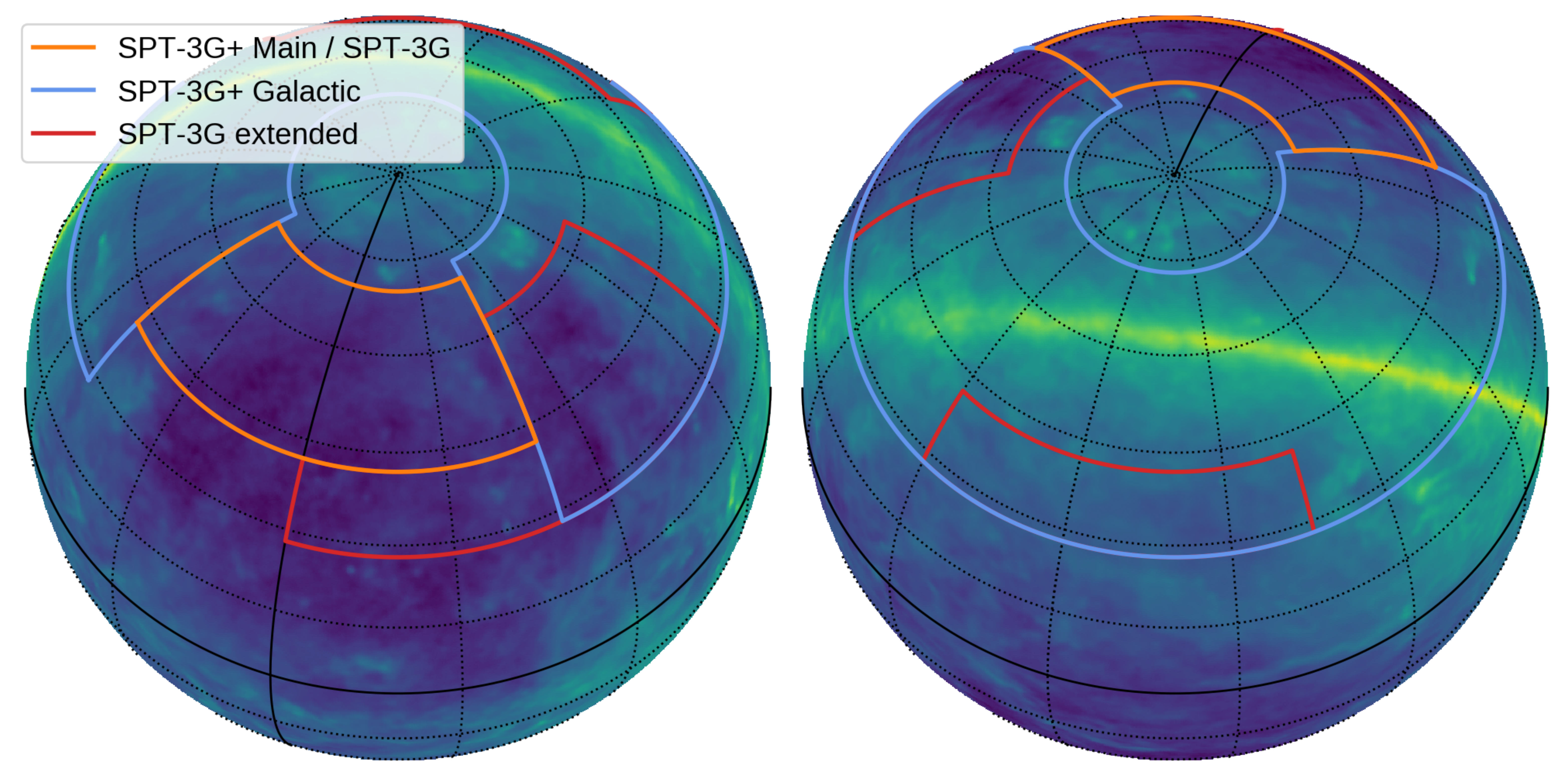}
\caption{Footprints of SPT-3G+ Main and Galactic surveys compared with SPT-3G survey and overlaid on the \planck\ thermal dust map~\cite{planck15-10}.}
\label{fig:surveyfootprint}
\end{figure}

\section{Site and Telescope Platform}
\label{sec:sitetelescope}

\begin{figure}[]
\centering
\includegraphics[width=0.6\textwidth]{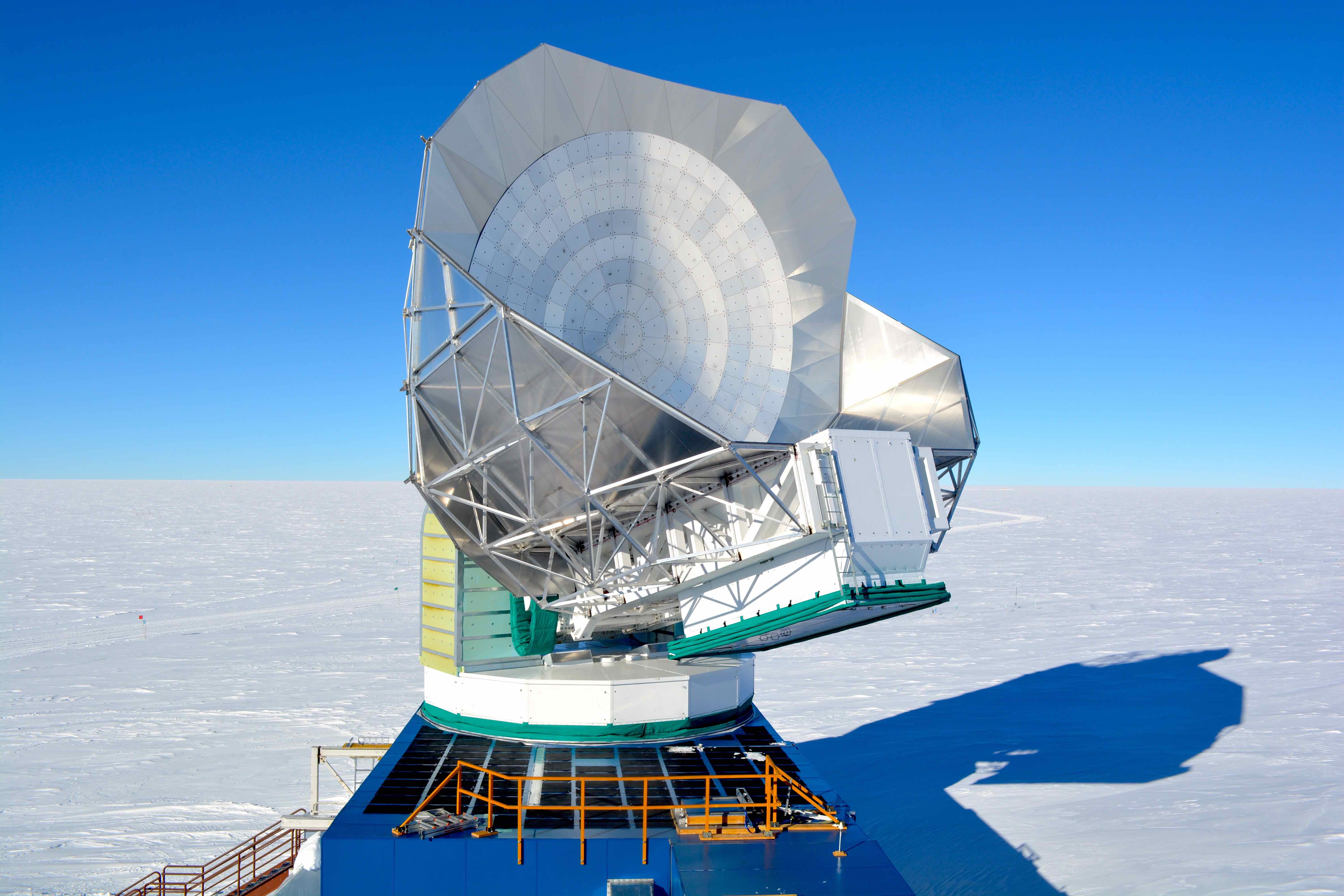}
\caption{
The 10-m aperture South Pole Telescope, observing during the 2017-2018 austral summer.
}
\label{fig:sptpicture}
\end{figure}

SPT-3G+ will make use of the 10-m aperture, sub-mm quality SPT, located at the geographic South Pole, the developed site with the world's best conditions for observing at mm- and submm-wavelengths (see \autoref{fig:sptpicture})~\cite{carlstrom11}.
The telescope and site are particularly well suited to the high-frequency CMB observations of SPT-3G+ for several reasons.
The primary mirror of SPT has a \SI{20}{\micro\meter} rms surface error, which is sufficiently high-quality to enable efficient observations at \SI{345}{\giga\hertz}.
The SPT-3G+ frequency bands have significant atmospheric absorption due to water vapor, and the median annual precipitable water vapor (pwv) at the South Pole is only \SI{0.32}{\milli\meter}, which helps enable high-frequency observations through nearly the entire austral winter~\cite{kuo17}.
The combination of low pwv, low atmospheric temperatures, absence of diurnal temperature variations, and laminar airflow produce uniquely stable atmospheric conditions that enable the measurement of large-scale cosmological modes, which is an important factor for the CMB anisotropy and Rayleigh scattering science goals.
In addition, observations taken during the austral summer season can still be successfully used in analyses of small-angular-scale phenomena; for example, summer data from the completed SPTpol survey were used to construct a catalog of galaxy clusters identified via the SZ effect~\cite{bleem20}.

\section{Instrument Design}
\label{sec:instrument}
To enable the science described in \autoref{sec:science}, the SPT-3G+ instrument reuses the primary and secondary optics of the SPT and includes a new 100~mK dilution-refrigerator-based cryostat housing low-loss silicon lenses and arrays of KIDs observing at 220, 285, and 345~GHz.
The design emphasizes efficient use of the optical throughput, with low-loss reimaging optics and high-density detector arrays, to make optimal use of the 2~deg$^2$ field of view of the SPT optics.
The use of modular optics tubes and modern, highly multiplexed RF readout electronics furthermore enables the cryostat to function as a platform for performing on-sky demonstrations of new KID-based detector technologies, including on-chip microwave spectrometers for line-intensity mapping.

\subsection{Cryostat}
\label{sec:cryostat}
The new detectors and lenses deployed by SPT-3G+ will be housed in a new cryostat that will replace the existing SPT-3G cryostat on the SPT.
This cryostat will contain a dilution refrigerator (DR) with the detectors operating at a base temperature of 100~mK.
The use of DRs in ground-based CMB experiments has been pioneered in recent years by projects including ACTPol~\cite{Thornton2016}, AdvACT~\cite{henderson16}, CLASS~\cite{Dahal2019}, and Simons Observatory~\cite{Zhu2021}.
Compared with the $^3$He-$^4$He sorption refrigerators that have historically been used by many projects, including SPT-3G~\cite{Sobrin2021}, DRs have several advantages: continuous operation, providing $\sim 20\%$ greater observing efficiency; lower base temperature, reducing generation-recombination noise at the detectors; and approximately $100\times$ greater cooling power at the detector operational temperature.

\begin{figure}[]
\centering
\includegraphics[width=0.99\textwidth]{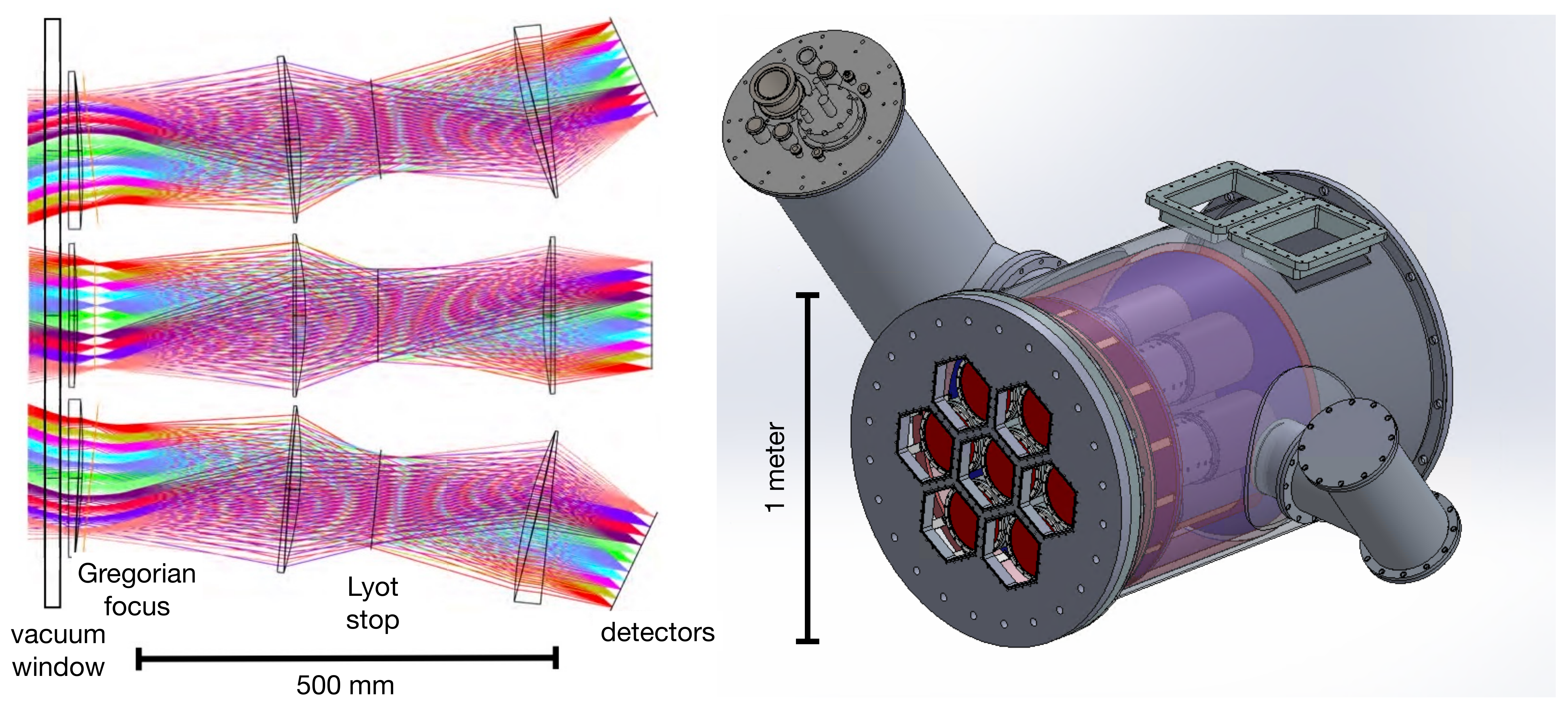}
\caption{\emph{Left:} Raytrace through a section of the SPT-3G+ cryogenic reimaging optics.
\emph{Right:} Rendering of the SPT-3G+ cryostat.
The secondary cylinder to the left side of the optics tubes contains a Bluefors SD-250 dilution refrigerator, while the secondary cylinder to the right can support a second pulse-tube cooler in addition to the one that cools the dilution refrigerator.}
\label{fig:cryooptics}
\end{figure}

The SPT-3G+ cryostat, shown in \autoref{fig:cryooptics}, will use a single Bluefors SD-250 DR\footnote{\url{https://bluefors.com/products/sd-dilution-refrigerator/}} to cool seven \SI{150}{\milli\meter} detector wafers to an operational temperature of \SI{100}{\milli\kelvin}.
The cryogenic reimaging optics (\autoref{sec:optics}) are housed in tubes and cooled to approximately \SI{4}{\kelvin}, with a design similar to the Simons Observatory~\cite{Zhu2021} and CMB-S4~\cite{Gallardo2022} large-aperture telescope cryostats.
A second CryoMech PT-415 pulse-tube cooler, in addition to the one integrated into the DR, provides extra cooling power at the \SI{4}{\kelvin} stage.
An ultra-high molecular weight polyethylene (UHMWPE) vacuum window and infrared filtering consisting of a combination of HD-30 closed cell foam sheets, alumina at \SI{40}{\kelvin}, and a metal-mesh low-pass filter at the \SI{4}{\kelvin} Lyot stop, complete the optics tube.

\subsection{Optics}
\label{sec:optics}
The cryogenic refracting optics of SPT-3G+ are split into seven independent tubes, which reimage sections of the Gregorian focus onto individual \SI{\sim135}{\milli\meter} diameter focal planes each with a \SI{0.7}{deg} diameter field of view with Strehl ratio $>0.9$ at \SI{375}{\giga\hertz}.
The single central tube uses a scaled-down copy of the SPT-3G optics design, with three rotationally symmetric, plano-convex, 6th-order asphere lenses~\cite{Sobrin2021}.
The six outer tubes use plano-convex lenses with non-rotationally symmetric surface sags, the design of which is optimized to correct coma in an off-axis section of the Gregorian focus.
Despite the more complex shape of the lenses in the outer optics tubes, the Strehl ratio and field of view of the outer tubes is similar to the central tube.
At the detector surface, the beams are telecentric in both on- and off-axis tubes.
Whereas the SPT-3G optics used \SI{720}{\milli\meter} diameter alumina lenses, SPT-3G+ will use \SI{200}{\milli\meter} Si lenses, and a Lyot stop with a temperature of \SI{4}{\kelvin} is located between the second and third lenses inside the vacuum window.
The stop restricts the illumination of the primary mirror to the inner \SI{9}{\meter} and defines the aperture efficiency of the beams of the feedhorns in the focal plane.

The multiple smaller optics tubes permits the use of silicon refractive optics, which both have lower bulk loss tangent than alternative materials and can be machined with meta-material anti-reflection coatings having very low reflectivity and wide bandwidth~\cite{Datta2013,Coughlin2018}.
For simplicity, although the SPT-3G+ detector arrays are all single-color (see \autoref{sec:cmbdetectors}), the \SI{220}{\giga\hertz} and \SI{285}{\giga\hertz} detectors will use the same broadband AR coating developed for the \SI{220}{\giga\hertz} and \SI{270}{\giga\hertz} dichroic arrays of SO and CMB-S4.
For \SI{345}{\giga\hertz}, single-band machined AR coatings in Si lenses have been demonstrated by the TolTEC experiment, and SPT-3G+ will implement a similar design~\cite{toltec2020}.

\subsection{CMB Detectors}
\label{sec:cmbdetectors}
\begin{table}[]
\renewcommand{\arraystretch}{1.2}
\setlength{\tabcolsep}{15pt}
\begin{center}
\begin{tabular}{l | c c c }
\hline\hline
Observing band     & 220 GHz  & 285 GHz  & 345 GHz  \\ \hline
Number of 150-\si{\milli\meter} wafers          & 2                              & 3                              & 2                              \\ % from Karia
Number of detectors       & \num{9744}                          & \num{14616}                         & \num{9744}                          \\ % from Karia
Number of readout lines             & 12                             & 18                             & 12                             \\ % from Karia / Adam
Fractional bandwidth      & 0.26                           & 0.20                           & 0.082                          \\ 
Pixel size [\si{\milli\meter}] ([$F\lambda$])  & 2.2 (1.25)                           & 2.2 (1.61)                           & 2.2 (1.95)                           \\ % from current prototypes
NET per detector [\si{\micro\kelvin_{\textrm{CMB}} \sqrt{\second}}]            & \num{540}         & \num{1300}          & \num{5700}                          \\ 
Camera NET [\si{\micro\kelvin_{\textrm{CMB}} \sqrt{\second}}]                & 6.2                            & 13                             & 65                            \\ % from NET per detector, divided by sqrt(Ndet)
Beam FWHM [arcmin] & 0.8                          & 0.6                            & 0.5                          \\ % from NSF MRI proposal
\hline
\end{tabular}
\caption{Detector parameters for the SPT-3G+ focal plane.}\label{tab:detector_params}
\end{center}
\end{table}

SPT-3G+ will deploy seven single-color arrays of feedhorn-coupled KIDs observing at bands centered at 220, 285, and 345~GHz with fractional bandwidths of 26\%, 20\%, and 8.2\% and 4,872 detectors per array.
The use of KIDs offers several critical advantages over TESs at frequencies above 200~GHz, including the ability to read out more detectors per detector array.
The need for per-detector wirebonds limits TES-based architectures to \num{\sim 2000} detectors per 150-mm silicon wafer.
With vastly fewer wirebonds required for each silicon wafer, the detector density of a KID array can be much higher than an equivalent array of TESs, and this results in significantly better mapping speed per unit of focal plane area~\cite{Barry2022}. 
For example, the mapping speed of an SPT-3G+ detector wafer is more than twice that of an equivalent detector wafer using TESs.

\begin{figure}[]
\centering
\includegraphics[width=0.8\textwidth]{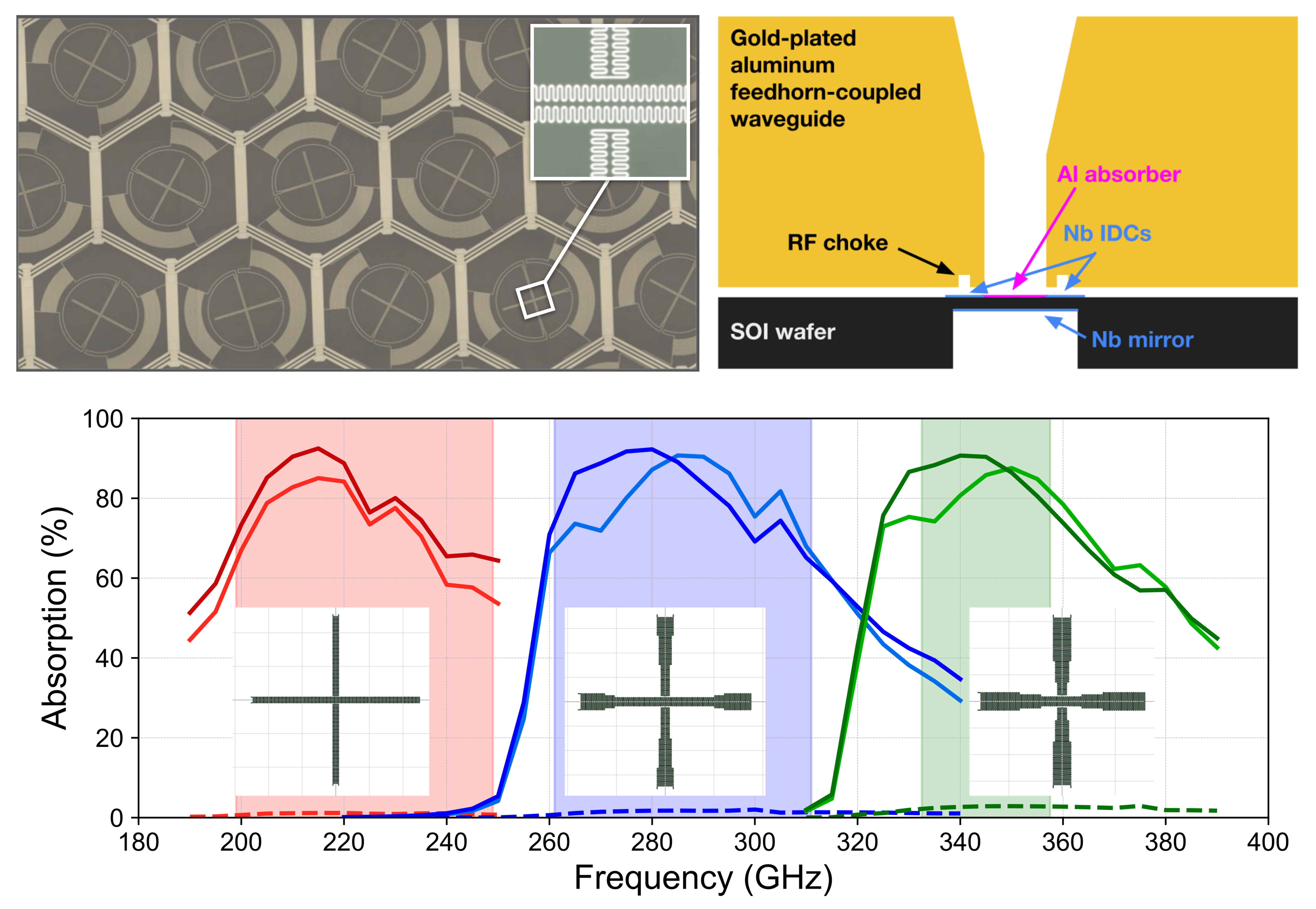}
\caption{Detector design and simulated performance.
\emph{Top left:} Tiling of \SI{220}{\giga\hertz} pixels in a detector array.
Inset shows the meander of the Al inductor, which constitutes the KID, at the junction between the two polarization arms of the direct-absorbing structure.
\emph{Top right:} Cross-sectional view of the feedhorn and waveguide that couple light from the telescope onto the pixel.
\emph{Bottom:} Simulated optical efficiency (solid) and cross-polarization (dashed) of the x and y detectors in each pixel.
Shaded regions represent the bandpass as defined by the waveguide cutoff of the feedhorn and the metal-mesh low-pass filter in the optics.
Insets show how the inductor design is scaled to have a different volume in each frequency band, accounting for the differences in optical loading.
Figure adapted from Dibert, \emph{et al.}~\cite{Dibert2021}.}
\label{fig:modulemontage}
\end{figure}

We have developed a direct-absorbing pixel design using an integrated backshort, which is simple to fabricate and achieves high optical efficiency and low cross-polarization coupling in simulations.
The pixel design, shown in \autoref{fig:modulemontage}, consist of two orthogonal meanders of Al that each form the inductor of a lumped-element KID~\cite{Dibert2021}.
Above the pixel is a gold-plated Al block that contains feedhorns that transition to waveguide.
Each of the KID meanders couples directly to one of the two orthogonal TE11 fundamental modes of the waveguide.
An RF choke around the opening of the waveguide suppresses optical crosstalk between adjacent pixels.
The alignment of the feedhorns to the pixels on the silicon wafer is defined by pin and slot alignment features removed from the silicon wafers using a deep reactive ion etch (DRIE), similar to the scheme used by the Simons Observatory~\cite{2021arXiv211201458M}.
The backside of the silicon detector wafer beneath each pixel is partially removed with a DRIE and then metallized with Nb in order to form $\lambda / 4$ backshort for the waveguide cavity.
The distance between the device layer and the backshort is controlled by using the insulator layer of a silicon-on-insulator wafer to stop the DRIE process at a well defined distance from the device layer.
The meander of the KID, shown in \autoref{fig:modulemontage}, serves to increase the volume of the inductor, increasing the quality factor of the resonator under optical load, while simultaneously maintaining low cross-polarization coupling.
Simulations indicate a loaded quality factor of $Q\sim 10^5$ for all frequency bands and a cross-polarization of $\lesssim 3$\%.
Since the pixels use direct-absorbing KIDs, the low-frequency edge of the bandpass is defined by the frequency cutoff of the waveguide, rather than an on-chip filter.
The high-frequency edge of the bandpass is set by a free-space, metal-mesh, low-pass filter, similar to those that are widely used in CMB experiments~\cite{2006SPIE.6275E..0UA}.

The KIDs are read out by coupling to Nb interdigitated capacitors and then capacitively coupled to a Nb coplanar waveguide (CPW), with pixels repeatedly patterned over a 150-mm-diameter wafer and split into six readout lines with a multiplexing factor of 812.
Achieving this multiplexing factor requires a fractional frequency placement precision of $\lesssim 10^{-4}$.
The IDCs of SPT-3G+ are designed to be compatible with laser trimming which allows for post-fabrication editing of the detector resonant frequencies and has been demonstrated to achieve this placement precision by several groups including our own~\cite{Shu:2018fex,McKenney:2018pds,McGeehan2018}.

%We have experimentally demonstrated many of the key performance metrics of this pixel architecture using test devices with 5-10 pixels at both 220~GHz and 345~GHz, and we have also scaled the design to 81-pixel subarrays utilizing a layout that can be tiled to cover one sixth of a 150-mm wafer.
% Ask Karia which data can be shared in this section. It is not clear to me.

% optical efficiency?
% noise peformance?
% FTS, bandpass
% crosspol
% Q-factor

%Array design and fabrication
Fabrication of test pixels and prototype subarrays has been performed at the Pritzker Nanofabrication Facility at the University of Chicago~\cite{Dibert2021}.
Following the initial maturation of design and fabrication techniques, the process has been transferred to the Center for Nanoscale Materials at Argonne National Laboratory, where the final detector arrays will be produced.
The pixel design entails a relatively simple fabrication process---with significantly fewer steps than fabrication of TES-based CMB detectors, for example---which has lent itself to a rapid design iteration cadence in response to lab testing.

%A 30nm thick layer of NbN is patterned on top of the Nb to prevent oxidation at the galvanic contact with the Al inductor.

\subsection{On-chip Spectrometers for Line Intensity Mapping}
\label{sec:limdetectors}
On-chip spectrometers using KIDs as the detector element in each channel are a scalable technology for LIM using mm-wavelength emission lines.
These spectrometers consist of an antenna or feedhorn-coupled orthomode transducer (OMT) which couples radiation from the telescope to a filter bank via microstrip.
A series of narrow-band filters ($\lambda / \Delta \lambda \sim 100-1000$) selects the radiation in each spectrometer channel, which is then coupled to a KID, where the microwave energy is dissipated as broken Cooper pairs and read out.
On-chip spectrometers have been developed by the DESHIMA~\cite{Endo2019} and SuperSpec~\cite{Karkare2020} collaborations, which are using them for galaxy spectroscopy.

\begin{figure}[]
\centering
\includegraphics[width=0.95\textwidth]{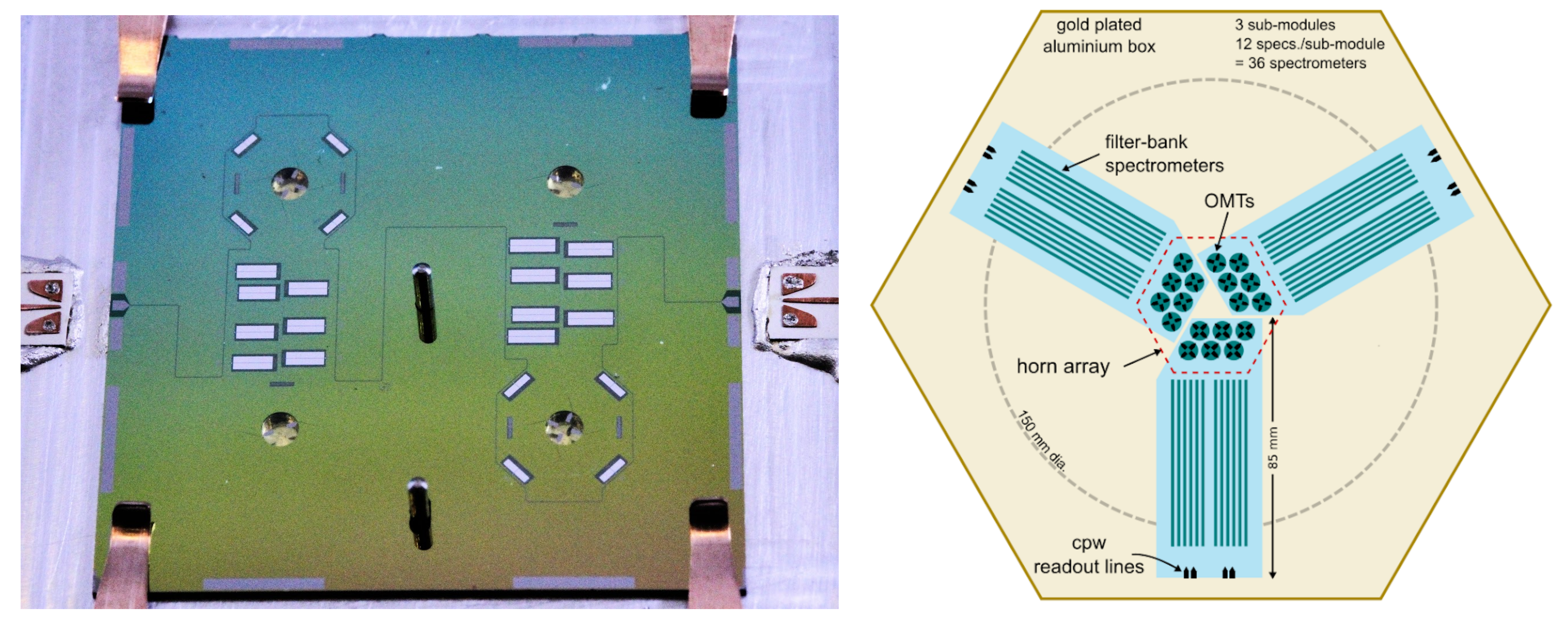}
\caption{
\emph{Left:} On-chip spectrometer test pixel fabricated at Argonne National Laboratory.
Optical power enters the system through an OMT and then is transmitted to a bank of eight filters that are read out by kinetic inductance detectors.
\emph{Right:} Concept for a focal plane of on-chip spectrometers being developed for the SPT-SLIM project. This focal plane design or a similar one with significantly more detectors could be deployed in one of the optics tubes of SPT-3G+.
Three separate silicon chips contain OMTs with filter banks arrayed radially around the central pixels.}
\label{fig:spectrometermontage}
\end{figure}

Several spectrometer technologies are currently being developed to demonstrate LIM targeting far-IR lines that are redshifted to the millimeter observing band, including gratings~\cite{TIME2014}, Fourier transform spectrometers (FTS)~\cite{Catalano2022}, and Fabry-Perot interferometers~\cite{Vavagiakis2018}, but on-chip spectrometers offer several advantages over these technologies.
All filtering and detector elements are fabricated on a single monolithic piece of silicon, which dramatically reduces the volume and mass of cryogenic optics.
In addition, the optical coupling to the telescope for an on-chip spectrometer can be identical to that of a CMB detector array, meaning that on-chip spectrometers are drop-in compatible with CMB experimental optics.
For this reason, the SPT-3G+ cryostat and readout are fully compatible with on-chip spectrometers, and we plan to deploy arrays in one or more of the optics tubes several years into the 4-year survey of SPT-3G+.

We are developing an on-chip spectrometer, closely following the SuperSpec design, with feedhorn-coupled OMTs, a prototype pixel of which is shown in \autoref{fig:spectrometermontage}.
The spectrometers observe in the atmospheric window between the \SI{118}{\giga\hertz} oxygen line and \SI{183}{\giga\hertz} water line, and a focal plane unit of these detectors will be demonstrated by the SPT-SLIM camera on the SPT during the 2023-2024 austral summer season~\cite{KarkareSLIM2022}.
Following the analysis of SPT-SLIM data, we will deploy either the same focal plane or an upgraded version in SPT-3G+.

\subsection{Readout Electronics}
The SPT-3G readout electronics consist of a room-temperature subsystem that handles digitization and synthesis of the resonator tones, and a cryogenic subsystem that consists of coaxial cabling and cryogenic low-noise amplifiers (LNAs) inside the cryostat. 
The room-temperature electronics are based on the ICE readout platform, which has successfully been used for readout of transition-edge sensors (TESs) in SPT-3G and in the signal processing for CHIME~\cite{bandura16}.
The ICE readout platform consists of a general-purpose digital motherboard containing a Xilinx Kintex-7 FPGA and an ARM processor, which couples to application-specific mezzanine daughter cards for digitization and synthesis.
This architecture affords significant flexibility for adapting to the readout requirements of different types of detectors, while maintaining a very high degree of hardware and software maturity
SPT-3G has taken data continuously using 32 IceBoards for 5 years, achieving background-limited white noise performance, excellent low-frequency stability, and negligible readout downtime, and much of the low-level control software framework for TES readout has already been adapted to function with KIDs~\cite{bender19}.

\begin{figure}[]
\centering
\includegraphics[width=0.8\textwidth]{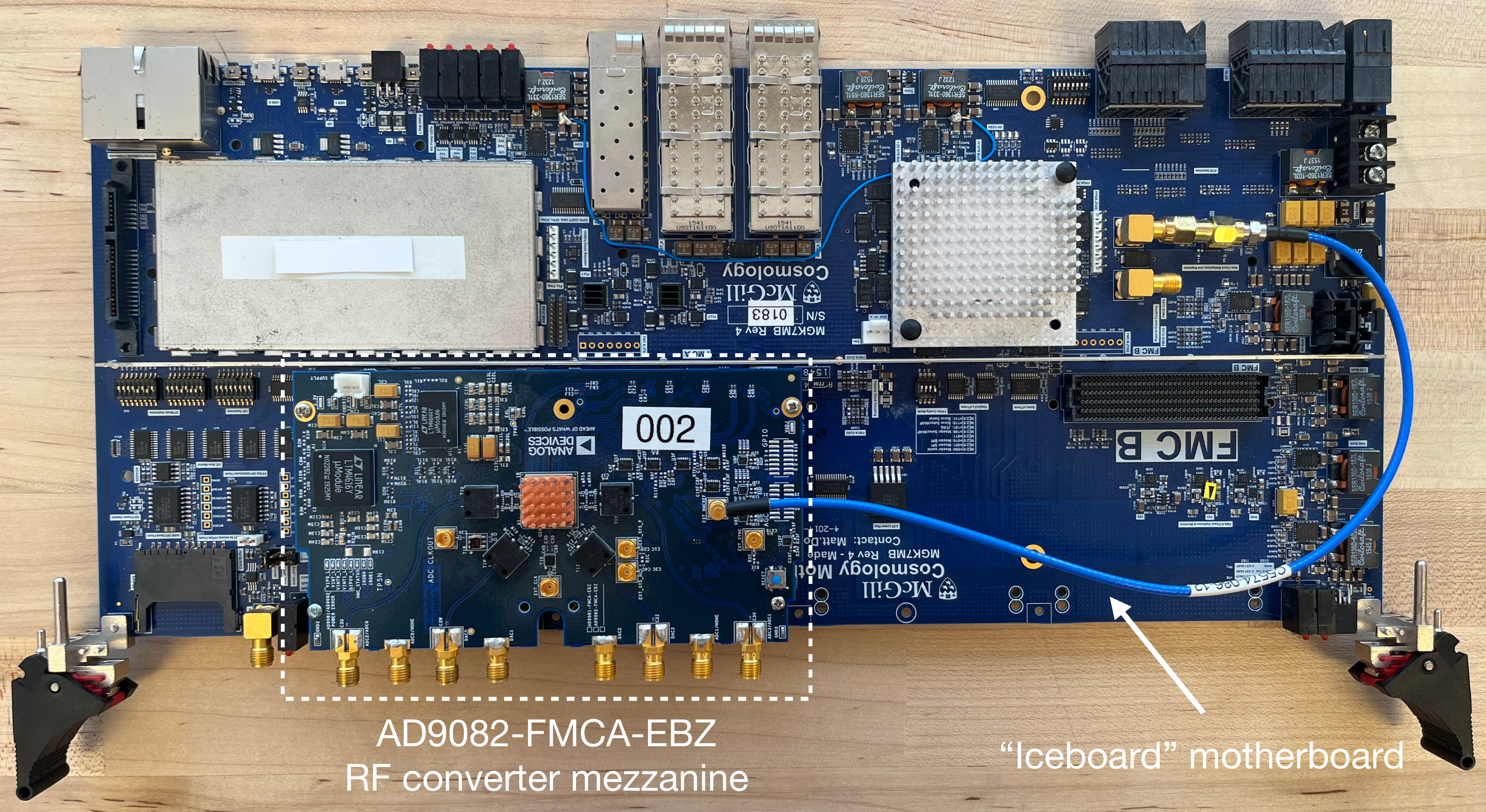}
\caption{
Digital ICE motherboard (``IceBoard'') with AD9082-FMCA-EBZ analog RF converter mezzanine board attached via a FPGA mezzanine connector (FMC).
The IceBoard contains an FPGA and an ARM processor, which handle the digital signal processing and streaming of data to the network and DAQ computer.
The DACs and ADCs on the RF converter mezzanine board handle the synthesis and digitization of probe tones.
Tones are generated directly at the detector resonant frequencies without the need for frontend analog mixers.
Figure adapted from Rouble, \emph{et al.} in these proceedings~\cite{Rouble2022}.
}
\label{fig:iceboard}
\end{figure}

SPT-3G+ has adapted the ICE platform to use with MKIDs by replacing the existing TES-style mezzanines with new GHz RF boards centered on the Analog Devices AD9082 device, which houses four 12 GSPS DACs and two 6 GSPS ADCs.
This RF frontend interrogates the KIDs at baseband, avoiding the need for mixers or other analog components except attenuators and amplifiers between the detectors and the warm electronics.
The firmware for TES readout used in SPT-3G~\cite{smecher2012} has been adapted to the faster ADCs and DACs needed for KIDs, implementing magnitude and phase measurement of the KID resonators at a fixed readout frequency, with a digital active feedback loop enabling continuous tone tracking if required.
% Could say something about nuller here? Probably not worth it, since we aren't using it.
A version of this firmware supporting $1024\times$ multiplexing per RF chain and using the AD9082-FMCA-EBZ evaluation board\footnote{\url{https://www.analog.com/en/design-center/evaluation-hardware-and-software/evaluation-boards-kits/eval-ad9082.html}} and has been successfully demonstrated in the lab with prototype \SI{220}{\giga\hertz} SPT-3G+ detectors.
More details on the electronics and firmware design and in-lab validation are given in Rouble, \emph{et al.} in these proceedings~\cite{Rouble2022}.

The cryogenic readout is a mixture of commercial off-the-shelf components similar to other mm/sub-mm KID-based cameras.
We plan to use stainless steel and CuNi coaxial cabling, with superconducting NbTi between the RF output side of the focal plane and the LNAs at the 4~K stage of the cryostat.
The design uses a single SiGe heterojunction bipolar transistor LNA per RF chain at the 4~K stage, which is produced by CryoElec\footnote{\url{https://www.cryoelec.com/}} and has +32~dB of gain at 1~GHz with a noise temperature of 3-4~K and a 1~dB compression point of -50~dBm.

\section{Conclusion}
Following the completion of SPT-3G survey operations, the SPT-3G+ camera will be deployed to the SPT, outfitted with KIDs to image the CMB in three bands centered at \SIlist[list-units=single]{220;285;345}{\giga\hertz}.
Together with the SPT-3G survey data, the SPT-3G+ dataset will yield constraints on the history of reionization, including the optical depth, with its clean measurements of the kSZ effect.
In addition, these data will provide a first hint of Rayleigh scattering at recombination, which could serve as a pathfinder for precision measurements from future space-based experiments.
At the same time, the survey data will enable the discovery of new high-redshift clusters and dusty star-forming galaxies.

The deep, high-frequency observations of the SPT-3G+ survey are enabled by dense focal planes with a total of \SI{34000} detectors.
These detectors use an efficient, direct-absorbing, feedhorn-coupled architecture of which prototypes have already been fabricated and tested in the \SIlist{220;345}{\giga\hertz} bands.
The leap in detector density is made possible by the RF-ICE readout platform, which has already achieved $1024\times$ multiplexing and inherits deployment-grade DAQ and control software from its SPT-3G heritage.
The modular optical design of SPT-3G+ has excellent image quality and would enable an eventual phased upgrade with on-chip spectrometers, expanding the science reach of the camera to include line-intensity mapping.
Finally, using the proven sub-millimeter-quality SPT maximizes the science impact of this hardware, without the need for new telescopes or ambient-temperature optics.

%\section{Deployment and Timeline}
%SPT-3G+ will deploy to the SPT during the 2024-2025 austral summer, when the current SPT-3G camera will finish its survey operations and be removed from the telescope.
%The initial deployment will include the full complement of detectors populating all seven optics tubes of the cryostat, and the four-year survey will begin operations during the 2025 austral winter season.
%Pending further development and demonstration of the on-chip spectrometer focal planes, including by the SPT-SLIM project, one or more of the optics tubes may be converted to use on-chip spectrometers in the second or third years of the survey.

\acknowledgments % equivalent to \section*{ACKNOWLEDGMENTS}       

The South Pole Telescope program is supported by the National Science Foundation (NSF) through the grant OPP-1852617.
Partial support is also provided by the Kavli Institute of Cosmological Physics at the University of Chicago.
Partial support for SPT-3G+ development is provided by NSF grant OPP-2117894.
This work made use of the Pritzker Nanofabrication Facility of the Institute for Molecular Engineering at the University of Chicago, which receives support from Soft and Hybrid Nanotechnology Experimental (SHyNE) Resource (NSF ECCS-2025633), a node of the National Science Foundation’s National Nanotechnology Coordinated Infrastructure.
Work supported by the Fermi National Accelerator Laboratory, managed and operated by Fermi Research Alliance, LLC under Contract No. DE-AC02-07CH11359 with the U.S. Department of Energy. 
Work at Argonne National Laboratory was supported by the U.S. Department of Energy (DOE), Office of Science, Office of High Energy Physics, under contract DE-AC02-06CH1137.
Work performed at the Center for Nanoscale Materials, a U.S. Department of Energy Office of Science User Facility, was supported by the U.S. DOE, Office of Basic Energy Sciences, under Contract No. DE-AC02-06CH11357.
ZP is supported by Argonne National Laboratory under award LDRD-2021-0186.
The McGill authors acknowledge funding from the Natural Sciences and Engineering Research Council of Canada and Canadian Institute for Advanced Research.
KD is supported by the Graduate Instrumentation Research Award through the Department of Energy, Office of High Energy Physics.
The Melbourne group acknowledges support from the Australian Research Council’s Discovery Projects scheme (DP210102386).
The U.S. Government retains and the publisher, by accepting the article for publication, acknowledges that the U.S. Government retains a non-exclusive, paid-up, irrevocable, world-wide license to publish or reproduce the published form of this manuscript, or allow others to do so, for U.S. Government purposes.

% References
\bibliography{spt} % bibliography data in report.bib
\bibliographystyle{spiebib} % makes bibtex use spiebib.bst

\end{document}